\documentclass[a4paper, 11pt] {article} 
\usepackage[utf8]{inputenc}
\usepackage[english]{babel}
\usepackage[left=2.cm, right=2.cm, bottom=2cm, top=2.cm]{geometry}
\usepackage{multicol, multirow, amsmath, amsthm, amssymb, amsfonts, fontenc, caption, graphicx, float, tocloft, booktabs, url, pdflscape, xcolor, physics, titlesec}
\usepackage[nottoc,numbib]{tocbibind} %Para que la bibliografía salga en el indice
\makeatletter
\def\@xfootnote[#1]{%
  \protected@xdef\@thefnmark{#1}%
  \@footnotemark\@footnotetext}
\makeatother

\makeatletter \renewcommand{\@cftmaketoctitle}{} \makeatother
\usepackage[colorlinks, linkcolor=blue, citecolor=blue, urlcolor=blue]{hyperref}
\numberwithin{equation}{section}

\titlespacing*{\subsubsection}
{5pt}{5pt}{1pt}

\begin{document}
\setlength{\abovedisplayskip}{6pt}
\setlength{\belowdisplayskip}{6pt}

\setlength{\intextsep}{0pt}

\begin{center}

{\Large{Revisiting the experimental test of Mermin's inequalities at IBMQ
}}

%\vspace{0.3cm}
%\large{Experimental implementation of Mermin's inequalities to probe into the hypothesis of local realism}

\vspace{5mm}
\normalsize{Diego González\textsuperscript{1}\footnote[*]{diegoj.gonzalez@estudiante.uam.es}, Diego Fernández de la Pradilla\textsuperscript{1}\footnote[$\dagger$]{diego.fernandezdelapradilla@estudiante.uam.es} and Guillermo González\textsuperscript{1}\footnote[$\ddagger$]{guillermo.gonzalezg01@estudiante.uam.es}}

\vspace{.2cm}
\textsuperscript{1} \textit{Instituto de Física Teórica, UAM-CSIC, Universidad Autónoma de Madrid, Cantoblanco, Madrid, Spain}
\end{center}

\begin{abstract}
Bell-type inequalities allow for experimental testing of local hidden variable theories. In the present work we show the violation of Mermin's inequalities in IBM's five-qubit quantum computers, ruling out the local realism hypothesis in quantum mechanics. Furthermore, our numerical results show significant improvement with respect to previous implementations. The circuit implementation of these inequalities is also proposed as a way of assessing the reliability of different quantum computers.
\end{abstract}
   
\hypersetup{linkcolor=black}
%\setcounter{tocdepth}{1} % Nota: Esto es para esconder subsections del índice. Se puede quitar si quereis.
%\tableofcontents

\hypersetup{linkcolor=blue}

\setcounter{page}{1}

\begin{multicols}{2}

\section{Introduction}\label{sec: intro}

In 1935, Einstein, Podolsky and Rosen (EPR) published a paper that challenged the consistency of the recently formulated quantum mechanics (QM) \cite{epr}. Their conclusion was that QM cannot be a complete theory, and they based their reasoning on the phenomenon of \textit{entanglement}. In other words, if reality observes local realism (LR), every element of reality has a well-defined value that cannot be modified in a superluminal way, and QM cannot be a complete theory. In its place, local hidden-variable theories (LHV) were proposed. 

For years, the difference between LHV and QM was essentially a metaphysical one, and it did not seem possible to empirically distinguish one from another. This changed radically in 1964 due to the contribution of the physicist John S. Bell \cite{bell}. In his paper, Bell proposed a set of physical quantities that could be measured and whose values, statistically, must satisfy some inequalities if local realism were valid. That is, if Bell inequalities hold, local realism prevails and QM is ruled out. When the experiments were performed the results were compatible with QM and contrary to the predictions of LHV \cite{clauser,aspect1,aspect2}. There are, actually, various experimental implementations of the usual two-particle Bell inequalities, notoriously the one proposed by Clauser, Horne, Shimony and Holt (CHSH) \cite{chsh}. Extended Bell-type experiments also allow for LR tests by using more than two entangled particles. Some examples are Mermin's and Svetlichny inequalities, which have been studied for 3 qubits and various quantum states in \cite{swain, sazim}.

In particular, Mermin's inequalities, proposed in 1990 \cite{mermin}, are one of the most significant examples of extended Bell-type inequalities to test nonlocal quantum correlations. The refutation of EPR provided by testing Mermin's inqualities is not intrinsically statistical; one single ideal measurement would suffice. However, the actual implementation in realistic quantum computers requires various thousands of shots to obtain statistical significance. The aim of this text is to implement Mermin's inequalities using IBM's five-qubit quantum computers \cite{ibm} through the \textit{IBM Quantum Experience} platform, and to verify that the results conflict with the classical bounds, for the cases with 3, 4 and 5 qubits. We will also compare our results with the existing literature. In particular, recent implementations include \cite{swain, alsina, german, 53}. 

The text is structured as follows: in section \ref{sec: mermin} we study in detail Mermin's inequalities and the quantum states that we need to prepare. We analyze the circuits to be implemented in section \ref{sec: circuitos}. In section \ref{sec: resultados} we gather the results and close with our conclusions in section \ref{sec: conclusiones}.

\section{Mermin's inequalities}\label{sec: mermin}

Mermin's inequalities can be implemented easily in a system with $n$ spins/qubits. GHZ \cite{ghz} states are particularly relevant, which are of the form
\begin{equation}
|\phi\rangle = \frac{1}{\sqrt{2}}\qty\big(|\underset{n \text{ qubits}}{\underline{0,0\cdots 0}}\rangle + e^{i\varphi}|\underset{n \text{ qubits}}{\underline{1,1\cdots 1}}\rangle ).
\end{equation}
We have
\begin{gather}
\sigma_z=\begin{pmatrix}
1 & 0 \\ 0 & -1
\end{pmatrix},\\
\sigma_x=\begin{pmatrix}
0 & 1 \\ 1 & 0
\end{pmatrix},\end{gather}
\begin{gather}
\sigma_y=\begin{pmatrix}
0 & -i \\ i & 0
\end{pmatrix},\\
|0\rangle=\begin{pmatrix}
1 \\ 0
\end{pmatrix},~ |1\rangle=\begin{pmatrix}
0 \\ 1
\end{pmatrix}.
\end{gather}
which implies that, for each qubit, the following holds
\begin{equation}
\sigma_z|0\rangle = |0\rangle \text{ y } \sigma_z|1\rangle = -|1\rangle.
\end{equation}
Starting from GHZ states, Mermin argues that the state
\begin{equation}
|\phi\rangle = \frac{1}{\sqrt{2}}\qty\big(|0,0\cdots 0\rangle + i| 1,1 \cdots 1 \rangle )
\end{equation}
is an eigenstate of the operator
\begin{equation}
M_n = \frac{1}{2i}\left[\bigotimes_{j=1}^n \left( \sigma_x^j+i\sigma_y^j\right) - \text{H.C.} \right],
\end{equation}
with eigenvalue $2^{n-1}$, where H.C. means hermitian conjugate. This is easy to check after realizing that
\begin{gather}
(\sigma_x+i\sigma_y)=2\sigma_+,\\
(\sigma_x-i\sigma_y)=2\sigma_-,\\
\sigma_+|0\rangle = 0;~~~~ \sigma_+|1\rangle = |0\rangle \text{ and}\\
\sigma_-|1\rangle = 0;~~~~ \sigma_-|0\rangle = |1\rangle.
\end{gather}
Next, Mermin expands the operator $M_n$. It is clear that only terms with an odd number of $\sigma_y$ survive when we substract the hermitian conjugate. Therefore, taking the expectation value with $|\phi\rangle$ yields
\begin{align}\nonumber
2^{n-1}=&\left\langle\sigma_{y}^{1} \sigma_{x}^{2} \cdots \sigma_{x}^{n}\right\rangle_{QM}+\cdots \\ \nonumber
&-\left\langle\sigma_{y}^{1} \sigma_{y}^{2} \sigma_{y}^{3} \sigma_{x}^{4} \cdots \sigma_{x}^{n}\right\rangle_{QM}-\cdots \\ \nonumber
&+\left\langle\sigma_{y}^{1} \cdots \sigma_{y}^{5} \sigma_{x}^{6} \cdots \sigma_{x}^{n}\right\rangle_{QM}+\cdots \\ \nonumber
&-\left\langle\sigma_{y}^{1} \cdots \sigma_{y}^{7} \sigma_{x}^{8} \cdots \sigma_{x}^{n}\right\rangle_{QM}-\cdots \\
&+\cdots = \langle M_n\rangle_{QM};
\end{align}
where the ellipses represent all the possible permutations of $\sigma_y$ in each row and the subscript QM means that this value has been computed within the framework of quantum mechanics.

The total number of terms is $2^{n-1}$, and each term is restricted to an interval with bounds $\pm1$. The conclusion is that $|\phi\rangle$ must be an eigenvector of each of the products of $\sigma_x$ and $\sigma_y$, which is also easy to check. Furthermore, the eigenvalue must be $(-1)^{(Y-1)/2}$, $Y$ being the number of $\sigma_y$ in each operator.

In this way, following Mermin's reasoning, we have found a combination of products of $\sigma_x$ and $\sigma_y$ that, if QM is valid, must be equal to $2^{n-1}$. Those combinations are generally referred to as \textit{Mermin polynomials}. Explicitly, those are:
\begin{align}
M_3 =&~\sigma_y^1\sigma_x^2\sigma_x^3 ~\underset{2\text{ more}}{\underline{+~ \cdots}} - \sigma_y^1\sigma_y^2\sigma_y^3,
\\ \nonumber
M_4 =&~\sigma_y^1\sigma_x^2\sigma_x^3\sigma_x^4 ~\underset{3\text{ more}}{\underline{+~\cdots}}\\ &- \sigma_y^1\sigma_y^2\sigma_y^3\sigma_x^4 ~\underset{3\text{ more}}{\underline{-~\cdots}}~~ \text{ and}
\\ \nonumber
M_5 =&~\sigma_y^1\sigma_x^2\sigma_x^3\sigma_x^4\sigma_x^5~\underset{4\text{ more}}{\underline{+~\cdots}}\\ \nonumber &- \sigma_y^1\sigma_y^2\sigma_y^3\sigma_x^4\sigma_x^5~\underset{9\text{ more}}{\underline{-~\cdots}}\\
&+ \sigma_y^1\sigma_y^2\sigma_y^3\sigma_y^4\sigma_y^5,
\end{align}
where the number below the ellipses indicates the number of terms with the same amount of $\sigma_y$ as the expectation value of the same row.

Even though we will not show it here \footnote{See \cite{mermin} for further details.}, a LHV theory that observes local realism predicts the values of the Mermin polynomials to be considerably lower than the ones obtained by using QM. More specifically,
\begin{align} \nonumber
&\langle M_n\rangle_{QM}=2^{n-1} \quad\text{and}\\ 
&\langle M_n\rangle_{LR}\leq \left\lbrace\begin{matrix}
2^{n/2} ~~~~~ n \text{ even and}\\
2^{(n-1)/2} ~~~~~ n \text{ odd}
\end{matrix}. \right.
\end{align}
It is precisely this disagreement what can be used to test the principle of local realism.

Until here we have reviewed Mermin's original paper. It is important to notice that the reasoning is based on the operator $M_n$ and the state $|\phi\rangle \propto |0,0\cdots 0\rangle + i |1,1\cdots 1\rangle$. If we modify the relative phase $\varphi$ between $|0,0\cdots 0\rangle$ and $|1,1\cdots 1\rangle$, the form of the associated Mermin polynomials changes too. This was the case with the expressions used by Alsina and Latorre \cite{alsina}. In particular, in their paper they generate Mermin polynomials from the recursive relation
\begin{equation}
M^A_{n}=\frac{1}{2}\left[ M^A_{n-1}\left(\sigma_x^{n}+\sigma_y^{n}\right)+ M_{n-1}^{A*}\left(\sigma_x^{n}-\sigma_y^n\right)\right]
\end{equation}
with ${M}^{A*}_n \equiv {M}_n(x\leftrightarrow y)$ and ${M}^A_1\equiv\sigma_x^1$. With this relation the obtained polynomials are:
\begin{align}
M^A_3 =&~M_3
\\ \nonumber
M^A_4 =&~-\sigma_x^1\sigma_x^2\sigma_x^3\sigma_x^4 - \sigma_y^1\sigma_y^2\sigma_y^3\sigma_y^4
\\ \nonumber
&~+ \sigma_y^1\sigma_x^2\sigma_x^3\sigma_x^4 ~\underset{3\text{ more}}{\underline{+~\cdots}}\\ \nonumber
&~- \sigma_y^1\sigma_y^2\sigma_y^3\sigma_x^4 ~\underset{3\text{ more}}{\underline{-~\cdots}} \\
&~+ \sigma_y^1\sigma_y^2\sigma_x^3\sigma_x^4 ~\underset{5\text{ more}}{\underline{+~\cdots}}~~ \text{ and}% \\ \nonumber
\end{align}
\begin{align}\nonumber
M^A_5 = &~-2 \sigma_x^1\sigma_x^2\sigma_x^3\sigma_x^4\sigma_x^5 \\ \nonumber
&~+ 2\qty\big( \sigma_x^1\sigma_x^2\sigma_x^3\sigma_y^4\sigma_y^5 ~\underset{9\text{ more}}{\underline{+~\cdots}}~)\\
&~- 2\qty\big( \sigma_x^1\sigma_y^2\sigma_y^3\sigma_y^4\sigma_y^5 ~\underset{4\text{ more}}{\underline{+~\cdots}}~).
\end{align}
Since any multiple of them can be used to check LR, they divided $M^A_5$ by 2.

In principle, all the qubits are independent and the state of the system should not be affected if we exchange any two qubits since they are all equivalent. Therefore,
\begin{align} \label{eq:1} \nonumber
\left\langle \sigma_y^1\sigma_x^2\cdots \sigma_x^n \right\rangle=\left\langle \sigma_x^1\sigma_y^2\sigma_x^3\cdots \sigma_x^n \right\rangle=\cdots \\ =\left\langle \sigma_x^1\cdots \sigma_x^{n-1}\sigma_y^n \right\rangle
\end{align}
and the same is true for every set of expectation values with the same number of $\sigma_x$ and $\sigma_y$. In this way we manage to reduce considerably the number of times we have to run the codes. As a verification, for the 3 qubits case we check this invariance experimentally.

Since the polynomials and states are different from the original proposal of Mermin \cite{mermin}, for the sake of completeness we also test the ones given in \cite{alsina}. We noticed the presence of two typos in \cite{alsina}, for the 4 and 5 qubit cases. In both of them, the relative phase in the GHZ state is incorrect by a factor of $-1$, which yields expectation values of the Mermin operators with a global $-1$ factor with respect to the ones found in the paper. Although it is not an important factor and the correction is trivial, it is here indicated so as to improve the replicability of these measurements.

To solve this conflict, we proceeded as follows. On the one hand, we took the GHZ states from \cite{alsina} and measured the primed polynomials
\begin{equation}
M_n^{A\prime} \equiv - M_n^A
\end{equation} 
referring to these results as {\color{purple}setup 1}. In addition to that, we prepared the GHZ states from \cite{alsina} with the corrected relative phase and we measured $\left\langle M_4^A \right\rangle$ and $\left\langle M_5^A \right\rangle$ ({\color{purple} setup 2}). Finally, Mermin's phase and polynomials were also used ({\color{purple} setup 3}). For $n=3$ there was no conflict and Mermin's and Alsinas' phases and polynomials are the same, so only one setup was used.

\section{Circuits}\label{sec: circuitos}

The circuit implementation consists in the creation of the state and the measurement of $\sigma^j_{x/y}$. The circuit that prepares the GHZ state is slightly different for each $n$ and $\varphi$. The changes needed for the generalization to an arbitrary number of qubits can be found by analyzing the simple patterns in the circuits shown here. Not all the qubits have an implemented CNOT gate in the computer, but in all the cases that we analyze we can exchange the role of the qubits and obtain one configuration where it can indeed be implemented.

In section \ref{sec: circuitosdibujitos} we show the various initial states, Mermin polynomials and circuits used for 3, 4 and 5 qubits. In section \ref{sec: circuitosmedida} we discuss how to measure the expectation values of each GHZ state. 

\subsection{GHZ state and polynomials}\label{sec: circuitosdibujitos}
\subsubsection*{Three qubits, only setup:}
\begin{gather}
|\phi_3\rangle = \frac{1}{\sqrt{2}}\left(|0,0,0\rangle + i| 1,1,1 \rangle \right)
\\
\langle M_3 \rangle = 3\langle\sigma_x\sigma_x\sigma_y\rangle-\langle\sigma_y\sigma_y\sigma_y\rangle
\\
\langle M_3 \rangle_{\mathrm{LR}}\leq 2
; ~~
\langle M_3 \rangle_{\mathrm{QM}}\leq 4
\end{gather}
\begin{figure}[H]
\centering
\includegraphics[height=0.9\linewidth, angle=270]{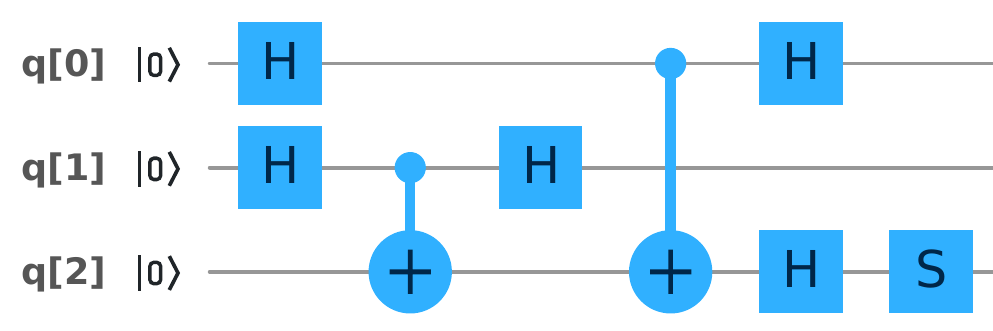}
\end{figure}
\subsubsection*{Four qubits, setup 1: A-L}
\begin{gather}
|\phi^{A}_4\rangle = \frac{1}{\sqrt{2}}\qty\big( |0,0,0,0\rangle + e^{-i\pi/4}|1,1,1,1\rangle )
\\ 
\begin{split}
\langle M_4^{A\prime} \rangle = 
\langle\sigma_x\sigma_x\sigma_x\sigma_x\rangle
-4\langle\sigma_x\sigma_x\sigma_x\sigma_y\rangle\\
-6\langle\sigma_x\sigma_x\sigma_y\sigma_y\rangle
+4\langle\sigma_x\sigma_y\sigma_y\sigma_y\rangle
+\langle\sigma_y\sigma_y\sigma_y\sigma_y\rangle
\end{split}
\\
\left\langle {M}^{A\prime}_4\right \rangle_{\mathrm{LR}}\leq 4
; ~~
\left\langle {M}^{A\prime}_4\right \rangle_{\mathrm{QM}}=8\sqrt{2}\approx 11.3
\end{gather}
\begin{figure}[H]
\centering
\includegraphics[height=0.9\linewidth, angle=270]{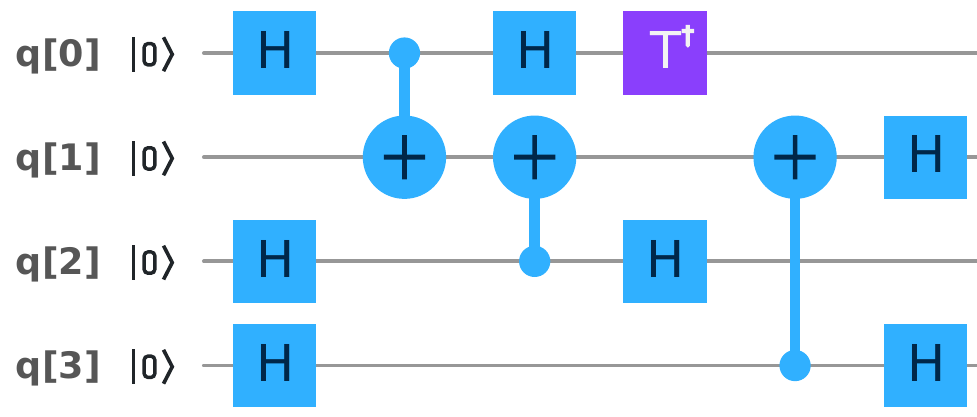}
\end{figure}
\subsubsection*{Four qubits, setup 2: Modified A-L}
\begin{gather}
|\phi^{A\prime}_4\rangle = \frac{1}{\sqrt{2}}\qty\big( |0,0,0,0\rangle - e^{-i\pi/4}|1,1,1,1\rangle )
\\
\begin{split}
\langle M_4^{A} \rangle = 
-\langle\sigma_x\sigma_x\sigma_x\sigma_x\rangle
+4\langle\sigma_x\sigma_x\sigma_x\sigma_y\rangle\\
+6\langle\sigma_x\sigma_x\sigma_y\sigma_y\rangle
-4\langle\sigma_x\sigma_y\sigma_y\sigma_y\rangle
-\langle\sigma_y\sigma_y\sigma_y\sigma_y\rangle
\end{split}
\\
\left\langle {M}^A_4\right \rangle_{\mathrm{LR}}\leq 4
; ~~
\left\langle {M}^A_4\right \rangle_{\mathrm{QM}}=8\sqrt{2}\approx 11.3
\end{gather}
\begin{figure}[H]
\centering
\includegraphics[height=0.9\linewidth, angle=270]{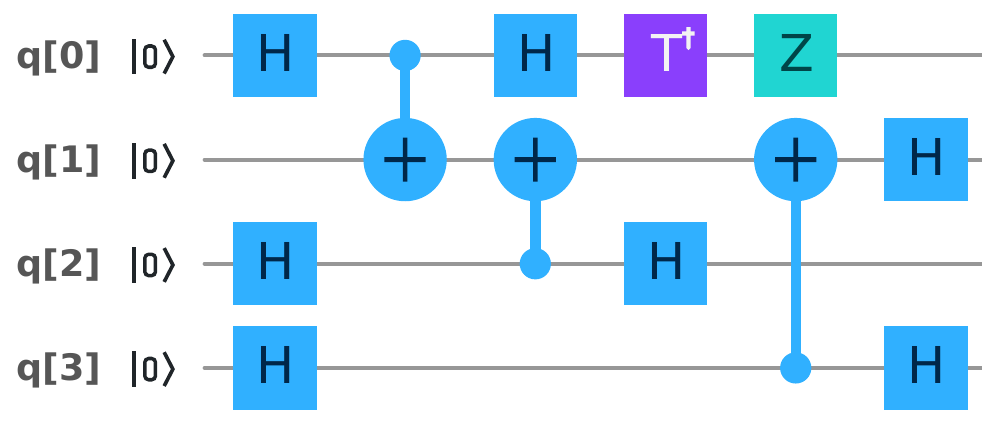}
\end{figure}
\subsubsection*{Four qubits, setup 3: Mermin}
\begin{gather}
|\phi_4\rangle = \frac{1}{\sqrt{2}}\qty\big(|0,0,0,0\rangle + i| 1,1,1,1 \rangle )
\\
\langle M_4 \rangle = 4\langle\sigma_x\sigma_x\sigma_x\sigma_y\rangle-4\langle\sigma_x\sigma_y\sigma_y\sigma_y\rangle
\\
\left\langle {M}_4\right \rangle_{\mathrm{LR}}\leq 4
; ~~
\left\langle {M}_4\right \rangle_{\mathrm{QM}}=8
\end{gather}
\begin{figure}[H]
\centering
\includegraphics[height=0.9\linewidth, angle=270]{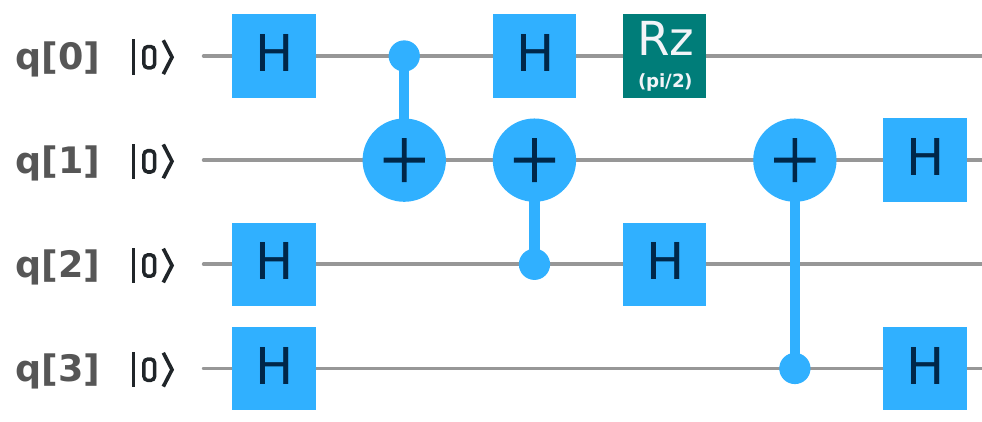}
\end{figure}

\subsubsection*{Five qubits, setup 1: A-L}
\begin{gather}
|\phi_5^A\rangle = \frac{1}{\sqrt{2}}\qty\big(|0,0,0,0,0\rangle + |1,1,1,1,1\rangle )
\\
\begin{split}
\langle M_5^{A\prime} \rangle =& \langle\sigma_x\sigma_x\sigma_x\sigma_x\sigma_x\rangle-10\langle\sigma_x\sigma_x\sigma_x\sigma_y\sigma_y\rangle\\&+5\langle\sigma_x\sigma_y\sigma_y\sigma_y\sigma_y\rangle
\end{split}
\\
\left\langle {M}^{A\prime}_5\right \rangle_{\mathrm{LR}}\leq 4
;~~
\left\langle {M}^{A\prime}_5\right \rangle_{\mathrm{QM}}=16
\end{gather}
\begin{figure}[H]
\centering
\includegraphics[height=\linewidth, angle=270]{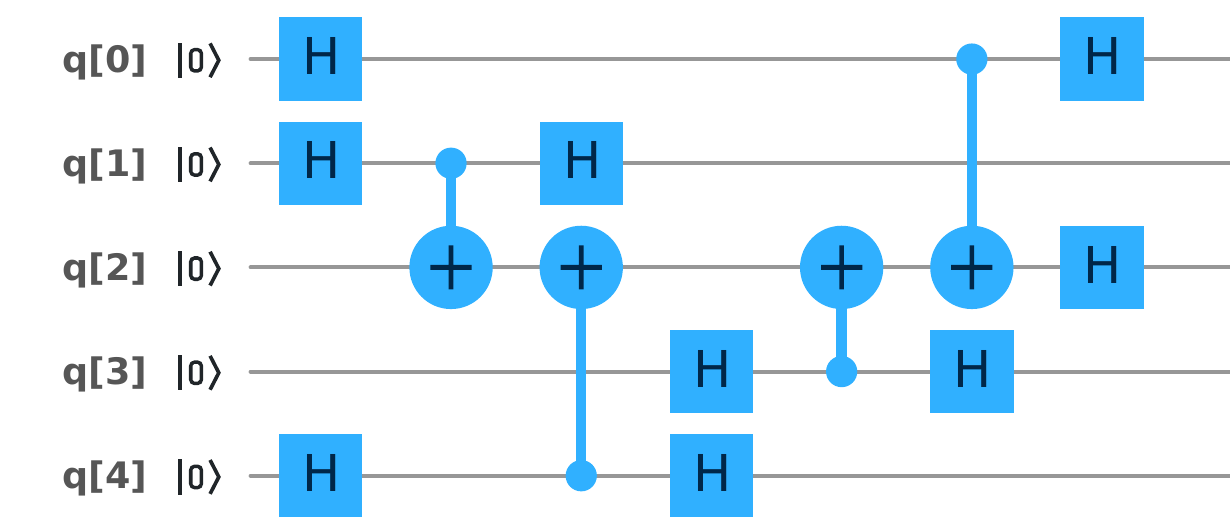}
\end{figure}
\subsubsection*{Five qubits, setup 2: Modified A-L}
\begin{gather}
|\phi_5^{A\prime}\rangle = \frac{1}{\sqrt{2}}\qty\big(|0,0,0,0,0\rangle - |1,1,1,1,1\rangle )
\\
\begin{split}
\langle M_5^{A} \rangle =& 
-\langle\sigma_x\sigma_x\sigma_x\sigma_x\sigma_x\rangle
+10\langle\sigma_x\sigma_x\sigma_x\sigma_y\sigma_y\rangle\\
&-5\langle\sigma_x\sigma_y\sigma_y\sigma_y\sigma_y\rangle
\end{split}
\\
\left\langle {M}^{A}_5\right \rangle_{\mathrm{LR}}\leq 4
;~~
\left\langle {M}^{A}_5\right \rangle_{\mathrm{QM}}=16
\end{gather}
\begin{figure}[H]
\centering
\includegraphics[height=\linewidth, angle=270]{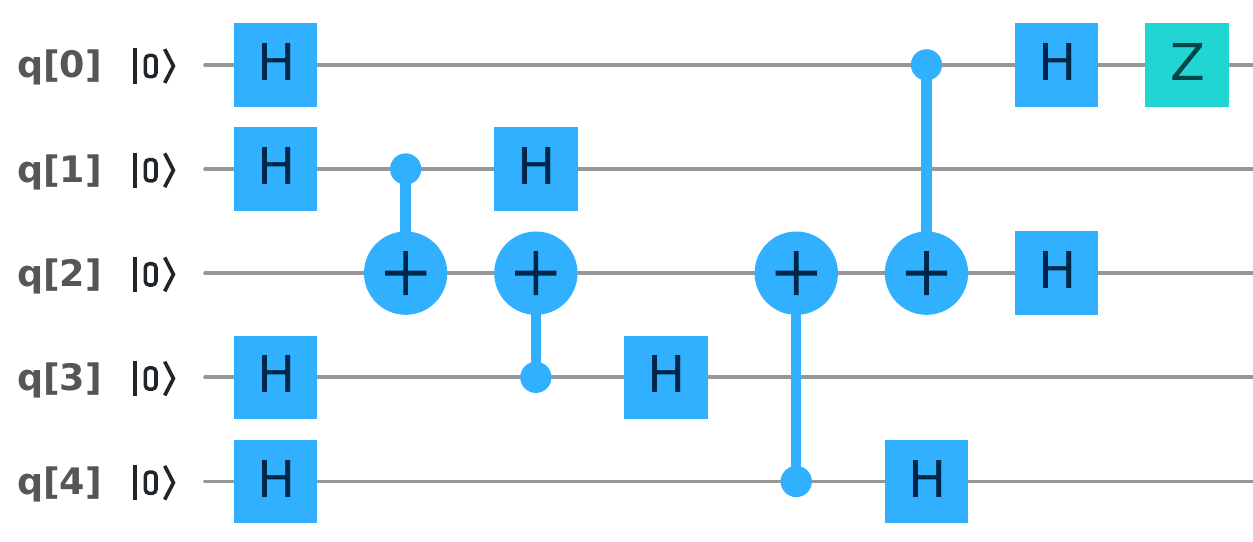}
\end{figure}
\vspace{5mm}
\subsubsection*{Five qubits, setup 3: Mermin}
\begin{gather}
|\phi_5\rangle = \frac{1}{\sqrt{2}}\qty\big(|0,0,0,0,0\rangle + i| 1,1,1,1,1 \rangle )
\\
\begin{split}
\langle M_5 \rangle =& 5\langle\sigma_x\sigma_x\sigma_x\sigma_x\sigma_y\rangle-10\langle\sigma_x\sigma_x\sigma_y\sigma_y\sigma_y\rangle\\&+\langle\sigma_y\sigma_y\sigma_y\sigma_y\sigma_y\rangle
\end{split}
\\
\left\langle {M}_5\right \rangle_{\mathrm{LR}}\leq 4
;~~
\left\langle {M}_5\right \rangle_{\mathrm{QM}}=16
\end{gather}
\begin{figure}[H]
\centering
\includegraphics[height=\linewidth, angle=270]{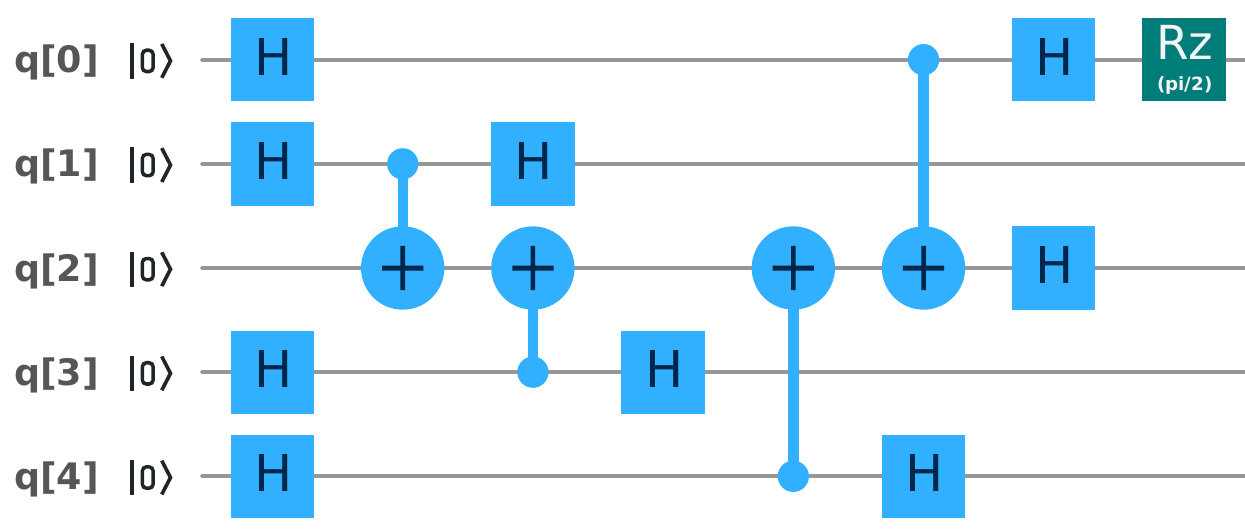}
\end{figure}

\subsection{Measurement} \label{sec: circuitosmedida}
\begin{figure}[H]
\centering
\includegraphics[height=0.8\linewidth, angle=270]{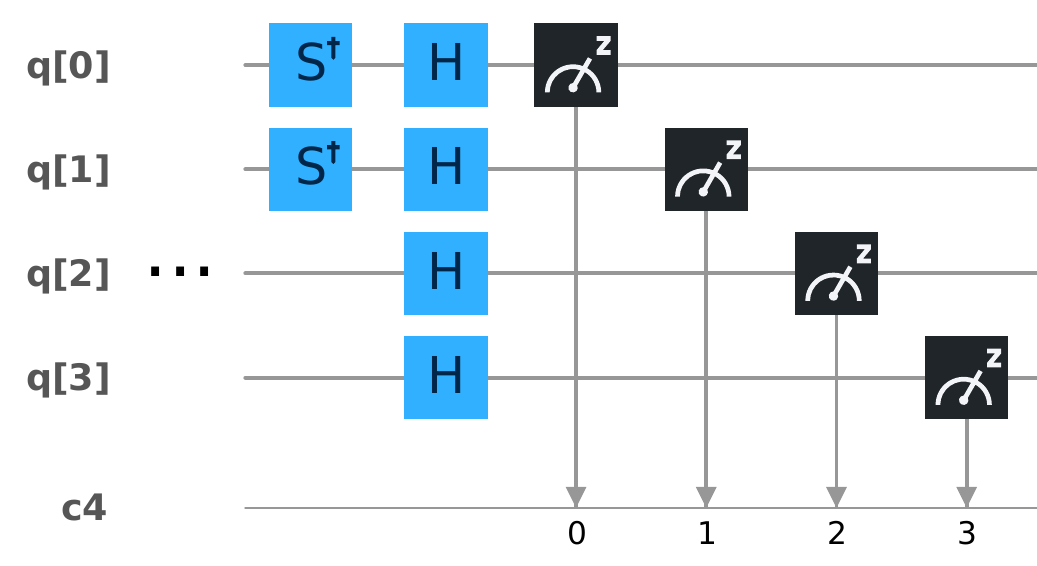}
\caption{Sample circuit that measures $\sigma^1_y\sigma^2_y\sigma^3_x\sigma^4_x$.}
\label{fig: medir}
\end{figure}
Finally, we cannot directly implement a measurement of $\sigma_x$ or $\sigma_y$ in the circuits. However, measuring $\sigma_{x(y)}$ directly is equivalent to acting on the qubit with $H (HS^\dag)$, and measuring $\sigma_z$ to acting with $S = \text{diag}(1,i)$ in the computational basis. We are going to explicitly show that this is the case for $\sigma_y$ here. A similar calculation can be done for $\sigma_x$. On one hand,
\begin{equation}
|0\rangle = \frac{1}{\sqrt{2}}(\underset{\text{Measuring }\sigma_y}{\underline{|+\rangle_y+|-\rangle_y}})\rightsquigarrow \frac{1}{\sqrt{2}}(\underset{\text{Measuring }\sigma_z}{\underline{|\hspace{0.5mm}0\hspace{0.5mm}\rangle+|\hspace{0.5mm}1\hspace{0.5mm}\rangle}})
\end{equation}
where the twisted arrows signals that we have replaced the eigenstates of $\sigma_y$ with eigenvalue 1 with the eigenstates of $\sigma_z$ with eigenvalue 1, and the same for the eigenstates with eigenvalue -1. In this way, measuring $\sigma_y$ over $|0\rangle$ and measuring $\sigma_z$ over the state after the arrow is equivalent. Furthermore,
\begin{equation}
|0\rangle \overset{S^\dag}{\longrightarrow} |0\rangle \overset{H}{\longrightarrow} \frac{1}{\sqrt{2}}(|0\rangle + |1\rangle),
\end{equation}
which is the same state that the previous operation of replacing $|+\rangle_y \rightsquigarrow |0\rangle$ and  $|-\rangle_y \rightsquigarrow |1\rangle$ lead to. The same happens with $|1\rangle$ as initial state. That is, measuring $\sigma_y$ is equivalent to acting with $HS^\dag$ and measuring $\sigma_z$. Statistically, the results will be identical.

\section{Experimental results}\label{sec: resultados}

In what follows we show the results obtained in our work. We have implemented the quantum circuits from section \ref{sec: circuitos} in the quantum computer prototypes from IBM through the \textit{IBM Quantum Experience}, in the cases of 3, 4 and 5 qubits. The circuits have been run in several computers, so as to compare the results from the different IBM machines. As for the uncertainties, section \ref{sec: error} contains the details on error propagation. We first present the general results in the next table.
\end{multicols}

{\centering
\begin{table}[H]
\caption{\textbf{Results of the present work}. The highest values of the Mermin polynomial for 3, 4 and 5 qubits for 16384 shots are shown, as well as local realism (LR) and quantum mechanics'(QM) upper bounds and the results of A-L\cite{alsina} and GM-S\cite{german}. All setups merged except for 4 qubit Mermin due to the different QM bound.}
\centering
\begin{tabular}{@{}cccccc@{}}
\toprule
					& LR & QM			& A-L			& GM-S			& \textbf{Present work} \\ 
\midrule
3 qubits				& 2  & 4				& $2.85\pm 0.02$	& $2.84\pm 0.07$	& \textbf{3.34} $\boldsymbol{\pm} $ \textbf{0.02}	\\
4 qubits				& 4  & 8$\sqrt{2}$	& $4.81\pm 0.06$	& $5.42\pm 0.04$	& \textbf{9.07} $\boldsymbol{\pm} $ \textbf{0.06}	\\
\textit{(Mermin)}	& 4  & 8				& -             	& -				& \textbf{6.14} $\boldsymbol{\pm} $ \textbf{0.04}	\\
5 qubits				& 4  & 16			& $4.05\pm 0.06$	& $7.06\pm 0.03$	&\textbf{10.33} $\boldsymbol{\pm} $ \textbf{0.08}	\\ 
\bottomrule
\end{tabular}
\end{table}}

\begin{multicols}{2}

In every case the measured expectation values are well above the threshold established by the local realism hypothesis, which is, hence, safe to be abandoned. 

We can see from the table that there has been a gradual improvement in the results as the IBM machines have been refined. The values from \cite{german} improved upon the values from \cite{alsina}, and so do ours upon the former. Most impressively, the case of 3 qubits, which barely changed from \cite{alsina} to \cite{german}, has increased greatly in our measurements. In the other cases, the results have gotten better in a more gradual fashion, clearly surpassing the bound imposed by local realism. Besides, the third row shows our results with the polynomial from \cite{mermin}, which also violates local realism. 

\end{multicols}

\subsection{3 qubits}
{\centering
\begin{table}[H]
\caption{\textbf{Results for 3 qubits.} We tested five different 5-qubit computers over 16384 shots each one. Expectation value for each individual circuit and result of the polynomial are shown with their respective errors.}
\label{tabla: 3qresult}
\centering
\begin{tabular}{@{}ccccc@{}}
\toprule
3 qubits	 & $\langle\sigma_x\sigma_x\sigma_y\rangle$	& $\langle\sigma_y\sigma_y\sigma_y\rangle$	& \textbf{Result}					\\
	& $\pm 0.007$	& $\pm 0.007$	& $\boldsymbol{\pm}$\textbf{0.02}	\\
\midrule
Vigo			& 0.835		& -0.744		& \textbf{3.25}  \\
Ourense		& 0.847		& -0.799		& \textbf{3.34}  \\
Valencia		& 0.662		& -0.607		& \textbf{2.59}  \\
Essex		& 0.690		& -0.494		& \textbf{2.56}  \\
IBMqx2		& 0.815		& -0.774		& \textbf{3.22}  \\
\bottomrule
\end{tabular}
\end{table}}

\begin{multicols}{2}

For all setups and qubit number the same five machines and number of shots have been used, to allow for a more systematic evaluation of the results. 

As can be seen in table \ref{tabla: 3qresult}, the values present some dependence on the particular IBM chip utilized. It is clear that although every single value in the bold column disagrees with local realism, the machines Vigo, Ourense and IBMqx2 provide much better results than Valencia and Essex.

For 3 qubits we also checked for the invariance under qubit exchange. As shown in table \ref{table: 3qexchange} we consistently found that $\langle\sigma_x\sigma_x\sigma_y\rangle\approx \langle\sigma_x\sigma_y\sigma_x\rangle\approx\langle\sigma_y\sigma_x\sigma_x\rangle$. More precisely, the standard deviation of those values was close enough to the experimental error that we could consider the hypothesis of invariance under qubit exchange \eqref{eq:1} to be experimentally confirmed and use this fact so as to greatly reduce the number of measurements to be carried out.

We may as well use the last column to test the accuracy of each machine as we did before. Again, the machines Vigo, Ourense and IBMqx2 have lower standard deviations among the three expectation values than the other two, Valencia and Essex, which is consistent with our previous comment.
\end{multicols}

{\centering
\begin{table}[H]
\caption{\textbf{Test of the invariance under qubit exchange.} 16384 shots were computed for all circuits and computers. Under perfect invariance, for each computer the standard deviation should be below the experimental error of each expectation value (in this case, $\pm 0.007$).}
\label{table: 3qexchange}
\centering
\begin{tabular}{@{}ccccc@{}}
\toprule
3 qubits   & $\langle\sigma_x\sigma_x\sigma_y\rangle$ & $\langle\sigma_x\sigma_y\sigma_x\rangle$ & $\langle\sigma_y\sigma_x\sigma_x\rangle$ & Standard \\
	& $\pm 0.007$	& $\pm 0.007$	& $\pm 0.007$	& deviation	\\
\midrule
Vigo		& 0.826		& 0.801		& 	0.812		& 0.012		\\
Ourense		& 0.847		& 0.797		& 	0.814		& 0.026		\\
Valencia	& 0.662		& 0.595		& 	0.651		& 0.036		\\
Essex		& 0.690		& 0.606		& 	0.618		& 0.045		\\
IBMqx2		& 0.815		& 0.789		& 	0.797		& 0.013		\\
\bottomrule
\end{tabular}
\end{table}}

\subsection{4 qubits}
{\centering
\begin{table}[H]
\caption{\textbf{Results for 4 qubits.} We tested five different 5-qubit computers over 16384 shots each one. Three different setups (of circuit and polynomial) were used, as described in section \ref{sec: circuitos}. Expectation values for each individual circuit and result of the polynomial are shown with their respective errors. Note that in the last table {\color{purple}(Setup 3)} the limits imposed by LR and QM are 4 and 8 respectively, while in the others they are 4 and $8\sqrt{2}$.}
\centering
\begin{tabular}{@{}ccccccc@{}}
\toprule
4 qubits & $\langle\sigma_x\sigma_x\sigma_x\sigma_x\rangle$ & $\langle\sigma_x\sigma_x\sigma_x\sigma_y\rangle$ & $\langle\sigma_x\sigma_x\sigma_y\sigma_y\rangle$ & $\langle\sigma_x\sigma_y\sigma_y\sigma_y\rangle$ & $\langle\sigma_y\sigma_y\sigma_y\sigma_y\rangle$ & \textbf{Result} \\
\color{purple}(Setup 1)  & $\pm 0.007$ & $\pm 0.007$ & $\pm 0.007$ & $\pm 0.007$ & $\pm 0.007$ & $\boldsymbol{\pm}$\textbf{0.06} \\
\midrule
Vigo		& 0.583		& -0.544		& -0.574		& 0.568		& 0.596		& \textbf{9.07}		\\
Ourense	& 0.608		& -0.511		& -0.579		& 0.493		& 0.549		& \textbf{8.65}		\\
Valencia	& 0.489		& -0.512		& -0.469		& 0.482		& 0.434		& \textbf{7.72}		\\
Essex	& 0.385		& -0.261		& -0.487		& 0.313		& 0.407		& \textbf{6.01}		\\
IBMqx2	& 0.407		& -0.199		& -0.450		& 0.216		& 0.401		& \textbf{5.17}		\\
\bottomrule
\end{tabular}
\end{table}}
~
{\centering
\begin{table}[H]
\centering
\begin{tabular}{@{}ccccccc@{}}
\toprule
4 qubits & $\langle\sigma_x\sigma_x\sigma_x\sigma_x\rangle$ & $\langle\sigma_x\sigma_x\sigma_x\sigma_y\rangle$ & $\langle\sigma_x\sigma_x\sigma_y\sigma_y\rangle$ & $\langle\sigma_x\sigma_y\sigma_y\sigma_y\rangle$ & $\langle\sigma_y\sigma_y\sigma_y\sigma_y\rangle$ & \textbf{Result} \\
\color{purple}(Setup 2)  & $\pm 0.007$ & $\pm 0.007$ & $\pm 0.007$ & $\pm 0.007$ & $\pm 0.007$ & $\boldsymbol{\pm}$\textbf{0.06} \\
\midrule
Vigo			& -0.521		& 0.616		& 0.527		& -0.590		& -0.497		& \textbf{9.00}		\\
Ourense		& -0.566		& 0.489		& 0.531		& -0.486		& -0.521		& \textbf{8.17}		\\
Valencia		& -0.480		& 0.543		& 0.465		& -0.573		& -0.410		& \textbf{8.14}		\\
Essex		& -0.522		& 0.314		& 0.502		& -0.305		& -0.438		& \textbf{6.45}		\\
IBMqx2		& -0.446		& 0.224		& 0.407		& -0.269		& -0.265		& \textbf{5.13}		\\
\bottomrule
\end{tabular}
\end{table}}

{\centering
\begin{table}[H]
\centering
\begin{tabular}{@{}cccc@{}}
\toprule
4 qubits & $\langle\sigma_x\sigma_x\sigma_x\sigma_y\rangle$ & $\langle\sigma_x\sigma_y\sigma_y\sigma_y\rangle$ & \textbf{Result} \\
\color{purple}(Setup 3) & $\pm 0.007$ & $\pm 0.007$ & $\boldsymbol{\pm}$\textbf{0.04} \\
\midrule
Vigo			& 0.776		& -0.759		& \textbf{6.14}		\\
Ourense		& 0.743		& -0.700		& \textbf{5.77}		\\
Valencia		& 0.625		& -0.660		& \textbf{5.14}		\\
Essex		& 0.522		& -0.508		& \textbf{4.12}		\\
IBMqx2		& 0.525		& -0.523		& \textbf{4.19}		\\
\bottomrule
\end{tabular}
\end{table}}

\begin{multicols}{2}

In the case of 4 qubits, all of the experimental results are above the bounds of local realism. From our results we conclude that both the polynomials from \cite{alsina} and \cite{mermin} can be employed to discard local realism. Futhermore, in the first two tables, every result compatible with $\langle M^A_4\rangle>8$  (Vigo, Ourense and Valencia) provides evidence for ``genuine four-particle non-locality'' as Alsina and Latorre put it \cite{alsina,Collins}.

It is interesting to note that the best results in these setups are those given by Vigo, Ourense and Valencia. In this regard, IBMqx2 falls off several ranking spots from the 3 qubit case, and Valencia joins the top three results.
\end{multicols}

\subsection{5 qubits}
{\centering
\begin{table}[H]
\caption{\textbf{Results for 5 qubits.} We tested five different 5-qubit computers over 16384 shots each one. Three different setups (of circuit and polynomial) were used, as described in section \ref{sec: circuitos}. Expectation value for each individual circuit and result of the polynomial are shown with their respective errors.}
\centering
\begin{tabular}{@{}ccccc@{}}
\toprule
5 qubits & $\langle\sigma_x\sigma_x\sigma_x\sigma_x\sigma_x\rangle$ & $\langle\sigma_x\sigma_x\sigma_x\sigma_y\sigma_y\rangle$ & $\langle\sigma_x\sigma_y\sigma_y\sigma_y\sigma_y\rangle$ & \textbf{Result} \\
\color{purple}(Setup 1)  & $\pm 0.008$ & $\pm 0.008$ & $\pm 0.008$ & $\boldsymbol{\pm}$\textbf{0.08} \\
\midrule
Vigo		& 0.719		& -0.589	& 0.656		& \textbf{9.89}		\\
Ourense		& 0.591		& -0.509	& 0.424		& \textbf{7.80}		\\
Valencia	& 0.517		& -0.420	& 0.455		& \textbf{6.99}		\\
Essex		& 0.506		& -0.413	& 0.220		& \textbf{5.74}		\\
IBMqx2		& 0.570		& -0.550	& 0.554		& \textbf{8.84}		\\
\bottomrule
\end{tabular}
\end{table}}
~
{\centering
\begin{table}[H]
\centering
\begin{tabular}{@{}ccccc@{}}
\toprule
5 qubits & $\langle\sigma_x\sigma_x\sigma_x\sigma_x\sigma_x\rangle$ & $\langle\sigma_x\sigma_x\sigma_x\sigma_y\sigma_y\rangle$ & $\langle\sigma_x\sigma_y\sigma_y\sigma_y\sigma_y\rangle$ & \textbf{Result} \\
\color{purple}(Setup 2)  & $\pm 0.008$ & $\pm 0.008$ & $\pm 0.008$ & $\boldsymbol{\pm}$\textbf{0.08} \\
\midrule
Vigo		& -0.683	& 0.611		& -0.552	& \textbf{9.56}		\\
Ourense		& -0.567	& 0.479		& -0.435	& \textbf{7.53}		\\
Valencia	& -0.587	& 0.440		& -0.424	& \textbf{7.11}		\\
Essex		& -0.470	& 0.366		& -0.469	& \textbf{6.47}		\\
IBMqx2		& -0.611	& 0.585		& -0.543	& \textbf{9.18}		\\
\bottomrule
\end{tabular}
\end{table}}
~
{\centering
\begin{table}[H]
\centering
\begin{tabular}{@{}ccccc@{}}
\toprule
5 qubits & $\langle\sigma_x\sigma_x\sigma_x\sigma_x\sigma_y\rangle$ & $\langle\sigma_x\sigma_x\sigma_y\sigma_y\sigma_y\rangle$ & $\langle\sigma_y\sigma_y\sigma_y\sigma_y\sigma_y\rangle$ & \textbf{Result} \\
\color{purple}(Setup 3)  & $\pm 0.008$ & $\pm 0.008$ & $\pm 0.008$ & $\boldsymbol{\pm}$\textbf{0.08} \\
\midrule
Vigo		& 0.711		& -0.622	& 0.554		& \textbf{10.33}	\\
Ourense		& 0.511		& -0.412	& 0.365		& \textbf{7.04}		\\
Valencia	& 0.515		& -0.468	& 0.412		& \textbf{7.66}		\\
Essex		& 0.385		& -0.344	& 0.371		& \textbf{5.74}		\\
IBMqx2		& 0.568		& -0.563	& 0.484		& \textbf{8.95}		\\
\bottomrule
\end{tabular}
\end{table}}

\begin{multicols}{2}

The 5 qubit case is analogous to the previous one. Every result contradicts the local realism hypothesis. In particular, the machines Vigo and IBMqx2 give excellent results, closely followed by Valencia and Ourense. Essex, on the other hand, falls short of the standard set by the other machines, although still well above the LR bound.

Again, our best results present a significant improvement with respect to the previous implementations of Mermin's inequalities. Our worst results are close to the results from \cite{german} which is a mere reflection of the fact that the IBM machines are not identical in terms of circuitry, nor are they all at the same stage of development.

\subsection{Error propagation} \label{sec: error}

In order to compute the final errors we proceed as follows. A Mermin polynomial is a linear combination of operators, which we can generically be written as
\begin{equation}
\bigotimes_{i=1}^n \sigma_{x/y}^i .
\end{equation}
When we measure on the quantum computer each of these products, there are $2^n$ possible outcomes, $(0,\cdots,0,0),~(0,\cdots,0,1),~(0,\cdots,1,0)$ and so on, each one with a given probability. We repeat the measurement 16384 times and count how many times each outcome occurs. Dividing over the total number of shots we obtain the probability that the outcome of the measurement of a particular operator. As for the uncertainty of the probability of each outcome, we follow A-L's approach \cite{alsina} and estimate it as
\begin{equation}
\delta p_{\text{outcome}} = \sqrt{\frac{p_{\text{outcome}}(1-p_{\text{outcome}})}{16384}}.
\end{equation}
The expectation value is 
\begin{equation}
\left\langle \bigotimes_{i=1}^n \sigma_{x/y}^i \right\rangle = \sum_{a,\cdots m,n = 0}^{1} p_{(a,\cdots,m,n)} (-1)^O,
\end{equation}
where $O$ is the number of $1$'s present in $(a,\cdots m,n)$. Hence, the error of the expectation value is
\begin{equation}
\delta \left\langle \bigotimes_{i=1}^n \sigma_{x/y}^i\right\rangle = \sqrt{\sum_{\text{outcome}}\left( \delta p_{\text{outcome}}\right)^2}.
\end{equation}
 
Finally, a given Mermin polynomial,
\begin{equation}
\langle M \rangle = \sum_{t} c_t\left(\left\langle\bigotimes_{i=1}^n \sigma_{x/y}^i \right\rangle\right)_t,
\end{equation}
has a propagated uncertainty given by
\begin{equation}
\delta \left\langle M\right\rangle = \sqrt{\sum_{t}c_t^2 \left(\delta \left\langle \bigotimes_{i=1}^n \sigma_{x/y}^i\right\rangle \right)_t^2}.
\end{equation}
The errors are rounded up to the first non-zero decimal place in the results, which is why they are equal or very similar among different columns and setups frequently.

\section{Conclusions}\label{sec: conclusiones}

We have experimentally verified the violation of Mermin's inequalities with 3, 4 and 5 qubits in the quantum computers of IBM. The results we have obtained allow us to reject the principle of local realism as a fundamental feature of reality, by an ample margin. Furthermore, addressing the comment by Alsina and Latorre \cite{alsina,Collins}, we have obtained $\langle M_4^{A}\rangle >8$, which implies that not only have we proved generic non-locality, but also genuine four-particle non-locality. In particular, we conclude that all of our results are incompatible with local realism.

On a less positive note, if we assume Quantum Mechanics to properly describe the results of, at least, these experiments, it is evident that the accumulated errors are still large. This is a major issue in quantum computing, and has given rise to a whole field of quantum error correction algorithms. Nevertheless, the tendency among our results and the previous attempts \cite{alsina,german} clearly indicates that the IBM computer prototypes are headed in the right direction.

In this same spirit, we also conclude that the violation of the Mermin inequalities can be used as a method to evaluate the reliability of quantum computers.

\section*{Acknowledgements}
We thank B. Zaldívar, A. Casas, E. López and G. Sierra for insightful conversations and advice.
We also thank the IBM Quantum team for making multiple devices available via the IBM Quantum Experience.
The access to the IBM Quantum Experience has been provided by the CSIC IBM Q Hub.

\section{Declarations}

\section*{Funding}

This work was done with financial support from the Universidad Autónoma de Madrid through the grant ``Ayudas de Investigación en Estudios de Máster''.

\section*{Conflicts of interest}

The authors declare that there is no conflict of interest.

\section*{Availability of data and material}

The data used for this study will be shared upon request.

\end{multicols}

\end{document}